\documentclass[pra,aps,10pt,twocolumn,amsmath,amssymb]{revtex4-2}

\usepackage[english]{babel}
\usepackage{graphicx}
\usepackage{bm}
\usepackage{color}
\usepackage{slashed}
\usepackage{latexsym}
\usepackage{appendix}
\newcommand{\dd}{d}
\newcommand{\ii}{i}
\newcommand{\ee}{e}

\newcommand{\me}{m_\mathrm{e}} 
\newcommand{\op}[1]{\hat{#1}}
\newcommand{\ket}[1]{|{#1}\rangle}
\newcommand{\bra}[1]{\langle{#1}|}
\newcommand{\braket}[2]{\langle{#1}|{#2}\rangle}

\begin{document}

\title{Nondipole signatures in ionization}
\author{M. C. Suster} \email{m.suster@uw.edu.pl}
\author{J. Derlikiewicz} \email{Julia.Derlikiewicz@fuw.edu.pl}
\author{K. Krajewska}
\author{F. Cajiao V\'elez} \email{Felipe.Cajiao-Velez@fuw.edu.pl}
\author{J. Z. Kami\'nski}

\affiliation{
Institute of Theoretical Physics, Faculty of Physics, University of Warsaw, Pasteura 5, 02-093 Warsaw, Poland }
\date{\today}

\begin{abstract}
A general method for solving numerically the time-dependent Schr\"odinger equation, that is based on the Suzuki-Trotter scheme with the split-step Fourier approach, is introduced. The method relies on a Hamiltonian decomposition, where except of the components depending exclusively on the momentum or on the position operators, there are also terms depending on both momentum and position operators in particular configurations. We demonstrate that, for as long as the latter does not depend on noncommuting coordinates of the momentum and position operators, nondipole effects in laser-matter interactions can be studied without applying extra unitary operations. Specifically, we analyze the effect of nondipole corrections in ionization of a two-dimensional hydrogen atom for low- and high-frequency pulses. In the former case, the electron wave packet dynamics is dominated by rescattering processes within the laser pulse, leading to the high-order harmonic generation. We illustrate that harmonics of even orders are generated in the direction of the laser field polarization. On the contrary, for high-frequency pulses, such rescattering processes can be neglected. We demonstrate that a significant portion of the low-energy photoelectrons is detected opposite to the laser pulse propagation direction as a consequence of their post-pulse wave packet spreading and interaction with the parent ion.
\end{abstract}

\maketitle

\section{Introduction}
\label{sec:introduction}

With the construction of a laser, scientists acquired an exceptional tool to generate coherent and intense beams of light. This source of 
radiation allows the observation of highly nonlinear phenomena such as above-threshold ionization (ATI)~\cite{Agostini1979} or high-order 
harmonic generation (HHG)~\cite{Franken1961}. In these processes, the electron interaction with the oscillating electromagnetic radiation 
is comparable in strength to the binding Coulomb forces inside the atom. Hence, the traditionally used perturbation theory is not applicable
to describe such processes. Novel theoretical approaches were therefore developed, including the strong-field approximation (SFA) in ATI \cite{sfa1,sfa2,sfa3}, 
the semiclassical three-steps model~\cite{Krause1992a,Krause1992b,Schafer1993,KulanderClassic,Corkum1993} or the Lewenstein model in 
HHG~\cite{Lewenstein1994}.  In the SFA the exact electron scattering state is approximated by the Volkov state in the laser 
field~\cite{Volkov1935}, i.e., the Coulomb interaction is neglected once the electron is promoted to the continuum. Such an approximation is fully 
justified in high-energy ionization of neutral atoms (when the Born approximation for the final scattering state can be applied) or in photodetachment 
from negative ions (as discussed in Refs.~\cite{GK,Pekin1,Pekin2,Pekin3}). However, at low photoelectron energies or when the initial target is 
a positively charged ion, the Coulomb interaction modifies the electron evolution importantly. Similarly, the Lewenstein model considers a free electron moving 
under the action of the electric field during its excursion to the continuum and before it recombines with the parent ion. It is the recombination 
process that leads to the emission of photons with energies which are integer multiples of the driving photon's energy, provided that the laser pulse 
is sufficiently long.

Due to the development of modern computers another approach, arising from the first principles of quantum mechanics, was established. This is by 
solving numerically the time-dependent Schr\"odinger equation (TDSE). It is typically based on the application of the finite-difference method and the 
Crank-Nicolson propagator while space and time variables are discretized~\cite{Grun1982}. Already in the 1980's this approach was used to analyze 
quantum effects in diverse scattering processes~\cite{Grun1982,Bottcher1982,Grun1983,Horbatsch1984}, to explore the ATI from hydrogen~\cite{Kulander1987}, 
or to study multiphoton ionization in a one-dimensional atomic model~\cite{Javanainen1988}. Other propagation methods include the Peaceman-Rachford method (see Refs.~\cite{PR55,ML65,Kono1997} and references therein) and the Suzuki-Trotter (ST) 
approach~\cite{Trotter,Suzuki1,Suzuki2,McLachlan,Muslu2005}, among others. Contrary to SFA, the numerical solution of TDSE does not offer analytical 
expressions for the probability amplitudes and requires a substantial computational effort, particularly when projecting the exact scattering 
state onto the field-free eigenstates of the atomic Hamiltonian. Nevertheless, TDSE offers an accurate description of quantum effects including 
rescattering, spreading of the electron wave packet, and Coulomb modifications of the electron trajectories in the continuum. For this reason, 
this ab-initio method is particularly useful in describing the low-energy ionization from atoms or positively-charged ions or in 
analyzing processes where the above mentioned quantum signatures are important.

In order to reduce the complexity of the above theoretical methods, the dipole approximation is typically used. In this approximation,
while the laser field varies in time, it is homogeneously distributed in space~\cite{Forre2014} (i.e., the laser field does not propagate). This 
is justified provided that the wavelength of the laser wave is considerably larger than the dimensions of atomic targets and so it breaks down for 
high-frequency fields. In addition, the influence of the magnetic component of the laser pulse on the electron dynamics has to be negligible 
as compared to that of the electric field. This, in turn, is satisfied for driving fields with low to moderate intensities. Note that in ionization 
by ultraintense laser pulses the magnetic field plays an important role in the evolution of the electron wave 
packet~\cite{Forre2005,Forre2006,Solve2009,Simonsen2015,Forre2016,Lindblom2018}. Also, the assumption of nonpropagating laser fields and the absence 
of a magnetic component render impossible to treat effects such as radiation pressure, where photons transfer momentum to the system in the direction 
of field propagation~\cite{Brennecke2018}. Furthermore, it has been shown, both experimentally and theoretically (see 
Refs.~\cite{Smeenk2011,Reiss2014,CKK2018} and references therein), that photoionization by near infrared laser pulses of moderate intensities already 
exhibits signatures of radiation pressure. This imposes a low-frequency limit on the applicability of the dipole approximation.

With the invention of the chirped-pulse amplification technique by Strickland and Mourou~\cite{Mourou1985} and with the development of free-electron lasers 
(FELs), extremely intense laser pulses with frequencies ranging from the x-ray regime down to the infrared regime can be obtained. For instance, 
in the HERCULES project, a 300TW Ti:Sapphire laser pulse is tightly focused such that intensity of $I\sim2\times10^{22}$~W/cm$^2$ can be 
achieved~\cite{Hercules,Yanovsky2008}. In addition, x-ray FELs (or XFELs) can produce coherent pulses of radiation with wavelengths down 
to $\lambda\sim0.1$~nm and intensities roughly $I\sim1\times10^{20}$~W/cm$^2$~\cite{Young2018}. Also, intense laser pulses 
in the long-wave infrared regime have been obtained, with wavelengths of the order of $\lambda\sim 9\mu$m~\cite{Wilson2019}. It is clear that laser-matter 
interactions in those high-(low-) frequency and high-intensity regimes are beyond the applicability of the dipole approximation.

Signatures of radiation pressure have been observed in ionization by high-intensity long-wavelength laser pulses. Such signatures appear as asymmetries 
in angular distributions of photoelectrons in the direction of laser field propagation~\cite{Ludwig2014,Lund2021,Smeenk2011,Maurer2018} 
(i.e., ionized electrons are detected with an additional momentum in that direction). Interestingly, under certain circumstances, a combination of Coulomb  
focusing and nondipole effects in elliptically polarized fields lead to a shift of the photoelectron distribution towards the laser 
source~\cite{Ludwig2014,Maurer2018,Danek2018} (see also Ref.~\cite{Peng2018}). Further, in photoionization by ultra-intense laser pulses important 
relativistic effects take place. For this reason, the electron dynamics should be described by the Dirac or Klein-Gordon equations, rather than by the  
Schr\"odinger equation. To this end, the relativistic strong-field approximation~\cite{Reiss1990} makes use of the relativistic Volkov 
solution~\cite{Volkov1935} while accounting for effects such as radiation pressure, spin dynamics, etc. It is also possible to introduce certain 
corrections into the expressions derived from the original SFA in order to account for relativistic effects. Such an approach is known as the 
quasi-relativistic SFA~\cite{KK2015,KCK2019}. Among those corrections one can mention: i) recoil or higher order Nordsieck 
corrections~\cite{Nordsieck1954}, which account for radiation pressure effects; ii) retardation correction, which reflects the fact that the laser field 
is a propagating wave; iii) relativistic mass correction, related to the relativistic variation of the electron mass; and others. Thus, it is possible 
to account for nondipole effects within the nonrelativistic framework. Other approaches, which have been successfully applied to analyze nondipole 
effects in both ATI~\cite{Boning2019,Boning2021,Boning2022} and HHG~\cite{Fritzsche2022}, consist in formulating an asymptotic solution of the Coulomb-free 
Schr\"odinger equation outside the dipole approximation. In such cases, the continuum states in the SFA (modified Volkov solutions) contain important 
nondipole signatures (see also the derivations in Refs.~\cite{Rosenberg1993,Rosenberg1994,Rosenberg2000,Gavrila2019,Boning2020} for electron scattering 
and atomic motion in multimode laser fields). 
Moreover, the nondipole dynamics in tunneling ionization has been recently studied in \cite{Keitel1,Keitel2}, and the strong-field ionization of atoms by an elliptically polarized laser field has been discussed in \cite{Milo2022}.
We note that nondipole effects have also been analyzed in few-photon ionization under the framework of the 
perturbation theory and the numerical solution of TDSE (see, e.g.,~\cite{Chelkowski2014,Peng2017,Peng2019}). Other theoretical approaches, including 
classical analysis in conjunction with the Monte Carlo method, have been thoroughly discussed in the review article \cite{Haram2020}.

The plan of this paper is as follows. In Sec.~\ref{sec:theory} we introduce the analytical framework together with the algorithm employed in our 
investigations. In Sec.~\ref{sec:atomic} we describe the two-dimensional Coulomb-like central potential used to model the target atom. 
While Sec.~\ref{sec:hsplitting} relates to the Hamiltonian decomposition for the Suzuki-Trotter method, both in the dipole and nondipole regimes, 
in Sec.~\ref{sec:laser} we define the laser pulse. The numerical parameters used for solving the time-dependent Schr\"odinger equation are 
presented in Sec.~\ref{sec:numeric}. The application of this model to describe physical phenomena is discussed in Sec.~\ref{sec:Mean-value}. By introducing 
the mean values of position, velocity, and acceleration of the electron wave packets we compare the electron dynamics with and without the dipole 
approximation. In particular, while in Sec.~\ref{sec:Low} the nondipole effects in HHG are explored for the low-frequency laser pulses, in 
Sec.~\ref{sec:High} we analyze the radiation pressure effects and forward drift of the electron wave packet in the high-frequency pulses. 
Sec.~\ref{sec:smd} concerns the backward electron propagation which, according to our investigations,  is a consequence of the post-pulse electron 
wave packet spreading and its interaction with the parent ion. Finally, in Sec.~\ref{sec:Conclusions} we present our conclusions.

In our numerical analysis, we use the atomic units of momentum $p_0=\alpha\me c$, energy $E_0=\alpha^2\me c^2$, length $a_0=\hbar/p_0$, time $t_0=\hbar/E_0$, and the electric field strength $\mathcal{E}_0=\alpha^3\me^2 c^3/(|e|\hbar)$, where $\me$ and $e=-|e|$ are the electron rest mass and charge, and $\alpha$ is the fine-structure constant. In analytical formulas we put $\hbar=1$, while keeping explicitly the remaining fundamental constants.

\section{Theoretical Background}
\label{sec:theory}

The dynamics of a quantum system is governed by the wave equation
\begin{equation}
\ii\partial_t\ket{\psi(t)}=\op{H}(\op{\bm{p}},\op{\bm{x}},t)\ket{\psi(t)},
\label{th1}
\end{equation}
where $\op{H}(\op{\bm{p}},\op{\bm{x}},t)$ is the Hamiltonian of the system under consideration, whereas $\op{\bm{p}}$ and $\op{\bm{x}}$ 
are the momentum and position operators, respectively. They fulfill the Heisenberg commutation relations,
\begin{equation}
[\op{x}_j,\op{p}_\ell]=\ii\delta_{j\ell},\quad j,\ell=1,\dots ,D,
\label{th2}
\end{equation}
where $D$ describes the spatial dimensionality of the system. Depending on the form of the Hamiltonian, Eq.~\eqref{th1} 
is called the Schr\"odinger, Pauli or Dirac equation.

The time-evolution of an initial state, $\ket{\psi(t_0)}$, is given by the unitary operator $\op{U}(t,t_0)$,
\begin{equation}
\ket{\psi(t)}=\op{U}(t,t_0)\ket{\psi(t_0)},
\label{th3}
\end{equation}
which satisfies an identical wave equation,
\begin{equation}
\ii\partial_t\op{U}(t,t_0)=\op{H}(\op{\bm{p}},\op{\bm{x}},t)\op{U}(t,t_0),
\label{th4}
\end{equation}
with the initial condition $\op{U}(t_0,t_0)=\op{I}$, where $\op{I}$ is the identity operator. The formal solution of this equation can be presented in the form,
\begin{equation}
\op{U}(t,t_0)=\op{\mathcal{T}}\Bigl[\exp\Bigl(-\ii\int_{t_0}^t \dd\tau \op{H}(\op{\bm{p}},\op{\bm{x}},\tau)\Bigr)\Bigr],
\label{th5}
\end{equation}
where $\op{\mathcal{T}}$ is the Dyson time-ordering operator. As follows from this expression, the evolution operator fulfills the decomposition relation,
\begin{equation}
\op{U}(t,t_0)=\op{U}(t,t')\op{U}(t',t_0),\quad t_0<t'<t,
\label{th6}
\end{equation}
which is the basis of the numerical integration of the wave equation \eqref{th1}. Indeed, by introducing the time discretization,
\begin{equation}
t_n=t_0+n\delta t, \quad n=0,1,\dots ,N, \quad \delta t=\frac{t-t_0}{N},
\label{th7}
\end{equation}
we obtain
\begin{equation}
\op{U}(t,t_0)=\op{U}(t_N,t_{N-1})\dots\op{U}(t_1,t_0)=\prod_{n=0}^{N-1}\op{U}(t_{n+1},t_n).
\label{th8}
\end{equation}
For a sufficiently small time increment $\delta t$ (or sufficiently large number of time steps $N$), we can use the approximation,
\begin{equation}
\op{U}(t_{n+1},t_n)=\ee^{-\ii\delta t \op{H}(\op{\bm{p}},\op{\bm{x}},\bar{t}_n)},
\label{th9}
\end{equation}
with some properly chosen $\bar{t}_n\in [t_n,t_{n+1}]$. 

Let us further assume that the Hamiltonian is the sum of two terms,
\begin{equation}
\op{H}(\op{\bm{p}},\op{\bm{x}},t)=\op{H}_1(\op{\bm{p}},\op{\bm{x}},t)+\op{h}_2(\op{\bm{p}},\op{\bm{x}},t).
\label{th10}
\end{equation}
Then, as follows from the analysis presented in Refs. \cite{Trotter,Suzuki1,Suzuki2}, we can approximate the evolution operator for infinitesimally small times, $\op{U}(t_{n+1},t_n)$, as
\begin{equation}
\op{U}(t_{n+1},t_n)=\op{U}_{ST2}(t_{n+1},t_n)+O((\delta t)^3),
\label{th11}
\end{equation}
where
\begin{align}
\op{U}_{ST2}(t_{n+1},t_n)=&\ee^{-\ii\frac{\delta t}{2} \op{H}_1(\op{\bm{p}},\op{\bm{x}},\bar{t}_n)}\ee^{-\ii\delta t \op{h}_2(\op{\bm{p}},\op{\bm{x}},\bar{t}_n)} \nonumber \\
 \times &\ee^{-\ii\frac{\delta t}{2}\op{H}_1(\op{\bm{p}},\op{\bm{x}},\bar{t}_n)}
\label{th12}
\end{align}
and $\bar{t}_n=(t_n+t_{n+1})/2$. Such an approximation is the essence of the Suzuki-Trotter method for the numerical solution of TDSE. 

Further, we assume that $\op{h}_2(\op{\bm{p}},\op{\bm{x}},t)$ also consists of two separate terms, i.e.,
\begin{equation}
\op{h}_2(\op{\bm{p}},\op{\bm{x}},t)=\op{H}_2(\op{\bm{p}},\op{\bm{x}},t)+\op{h}_3(\op{\bm{p}},\op{\bm{x}},t).
\label{th13}
\end{equation}
In this case, and according to the Suzuki-Trotter method, the infinitesimal evolution operator reads
\begin{equation}
\op{U}(t_{n+1},t_n)=\op{U}_{ST3}(t_{n+1},t_n)+O((\delta t)^3),
\label{th14}
\end{equation}
with
\begin{align}
\op{U}_{ST3}(t_{n+1},t_n)=&\ee^{-\ii\frac{\delta t}{2} \op{H}_1(\op{\bm{p}},\op{\bm{x}},\bar{t}_n)}\ee^{-\ii\frac{\delta t}{2} \op{H}_2(\op{\bm{p}},\op{\bm{x}},\bar{t}_n)} \nonumber \\
 \times &\ee^{-\ii\delta t \op{h}_3(\op{\bm{p}},\op{\bm{x}},\bar{t}_n)}\ee^{-\ii\frac{\delta t}{2}\op{H}_2(\op{\bm{p}},\op{\bm{x}},\bar{t}_n)}\nonumber \\
 \times &\ee^{-\ii\frac{\delta t}{2}\op{H}_1(\op{\bm{p}},\op{\bm{x}},\bar{t}_n)}.
\label{th15}
\end{align}
If desired, such a procedure can be sequentially repeated $\mathcal{K}$ times in order to account for $\mathcal{K}$
decompositions of the initial Hamiltonian. To identify the number of those decompositions, in Eqs.~\eqref{th11} to~\eqref{th15} we have used 
the subscript $ST\mathcal{K}$, with $\mathcal{K}=2\textrm{ or }3$ and $ST$ indicating that the Suzuki-Trotter approximation is used. 
Finally, from Eq.~\eqref{th8} we arrive at an expression for the evolution operator with the Hamiltonian decomposed 
into $\mathcal{K}$-parts,
\begin{equation}
\op{U}(t,t_0)=\prod_{n=0}^{N-1}\op{U}_{ST\mathcal{K}}(t_{n+1},t_n)+O((\delta t)^2).
\label{th16}
\end{equation}
Note that the overall error of this method is of the order of $(\delta t)^2$. It can be made smaller by applying 
additional corrections~\cite{Suzuki2,McLachlan,Hairer,Skokos}, in particular those proposed by Suzuki~\cite{Suzuki3} and 
Yoshida~\cite{Yoshida}. Here, it is also important to note that the approximations introduced above preserve the unitarity of the evolution 
operator, provided that the Hamiltonian is Hermitian.

\begin{figure}
  \includegraphics[width=0.8\linewidth]{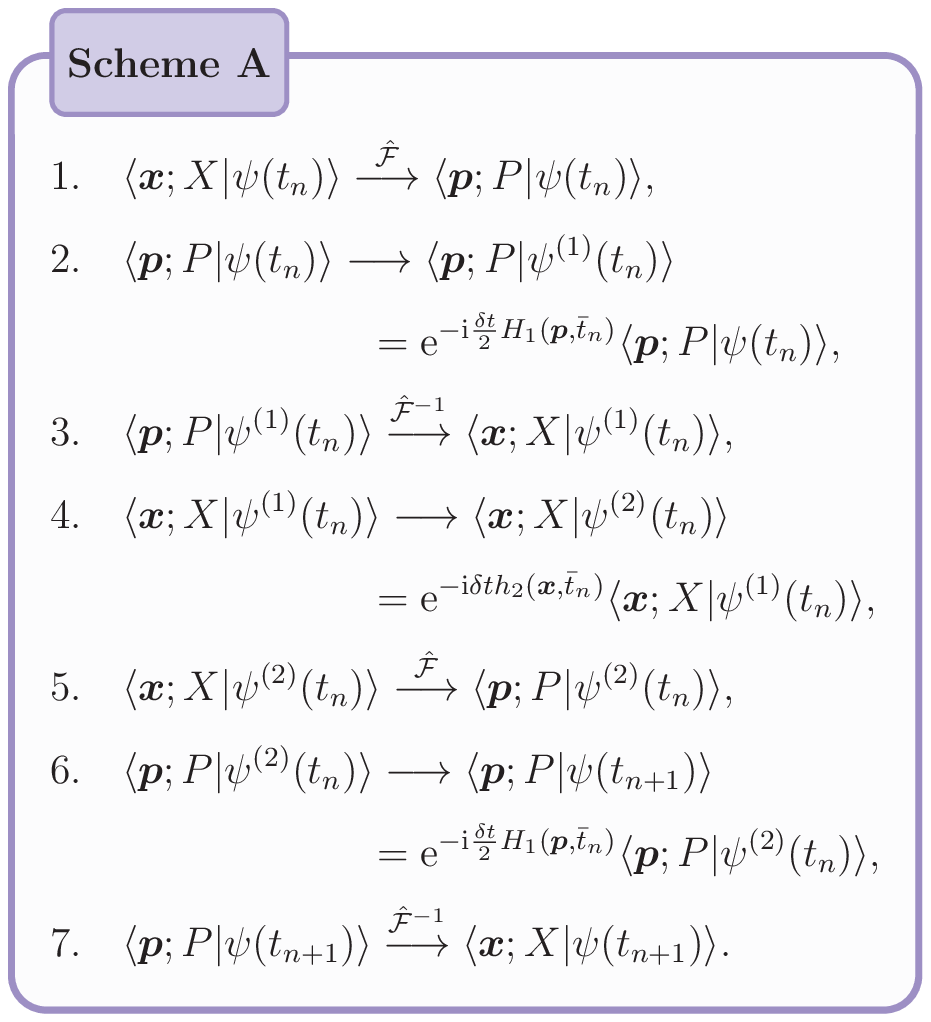}
\caption{Scheme A for the Suzuki-Trotter split-step Fourier method with a Hamiltonian divided into two terms [see, Eq.~\eqref{th12}]. 
It is assumed that each Hamiltonian component depends either on the momentum or on the position operator [Eq.~\eqref{th22a}], 
and that the wave function describing the quantum state of the system at time $t_n$, $\ket{\psi(t_n)}$, is known. In this scheme, 
we illustrate the steps that lead to $\ket{\psi(t_{n+1})}$ in position representation. Note that this type of algorithm is commonly used within the dipole approximation.
}
\label{SchemeAa}
\end{figure}

The power of the Suzuki-Trotter approximation is that the action of each exponent operator can be efficiently 
applied over any state of the system. For this to happen, the Hamiltonian decompositions should have some particular properties. For instance, if in Eq.~\eqref{th10} the operators $\op{H}_1$ and $\op{h}_2$ depend only on momentum and position operators, respectively, the Fast Fourier Transform algorithm~\cite{FFT1,FFT2,FFT3} can be applied. In order to proceed further, let us denote by $\ket{\bm{x};X}$ and $\ket{\bm{p};P}$ the eigenstates of the position and momentum operators,
\begin{equation}
\op{\bm{x}}\ket{\bm{x};X}=\bm{x}\ket{\bm{x};X}, \quad \op{\bm{p}}\ket{\bm{p};P}=\bm{p}\ket{\bm{p};P},
\label{th17}
\end{equation}
which, in a $D$-dimensional space, satisfy the orthogonality and completeness relations,
\begin{align}
\braket{\bm{x}';X}{\bm{x};X}=&\delta^{(D)}(\bm{x}-\bm{x}'), \nonumber \\
\braket{\bm{p}';P}{\bm{p};P}=&(2\pi)^D\delta^{(D)}(\bm{p}-\bm{p}'),
\label{th18}
\end{align}
\begin{equation}
\op{I}=\int\dd^Dx\ket{\bm{x};X}\bra{\bm{x};X}=\int\frac{\dd^Dp}{(2\pi)^D}\ket{\bm{p};P}\bra{\bm{p};P},
\label{th19}
\end{equation}
together with
\begin{equation}
\braket{\bm{x};X}{\bm{p};P}=\ee^{\ii\bm{p}\cdot\bm{x}}.
\label{th19b}
\end{equation}
Therefore, for any state $\ket{\phi}$ we can write
\begin{equation}
\bra{\bm{x};X}\ee^{-\ii\delta t\op{h}_2(\op{\bm{x}},t)}\ket{\phi}=\ee^{-\ii\delta t {h}_2({\bm{x}},t)}\braket{\bm{x};X}{\phi}
\label{th20}
\end{equation}
and
\begin{equation}
\bra{\bm{p};P}\ee^{-\ii\delta t\op{H}_1(\op{\bm{p}},t)}\ket{\phi}=\ee^{-\ii\delta t {H}_1({\bm{p}},t)}\braket{\bm{p};P}{\phi},
\label{th21}
\end{equation}
where the functions $\braket{\bm{x};X}{\phi}$ and $\braket{\bm{p};P}{\phi}$ are related through the Fourier transform,
\begin{equation}
\braket{\bm{p};P}{\phi}=\int\dd^Dx\ee^{-\ii\bm{p}\cdot\bm{x}}\braket{\bm{x};X}{\phi}.
\label{th22}
\end{equation}
Hence, we can determine the evolution operator given by Eqs.~\eqref{th11} and~\eqref{th12}. In order to illustrate this,  
in Fig.~\ref{SchemeAa} we show the steps for the temporal evolution of a system according to the Suzuki-Trotter split-step 
Fourier method (Scheme A). It is assumed that such a system is originally found in the state $\ket{\psi(t_n)}$ and that the Hamiltonian can be decomposed into two parts ($\mathcal{K}=2$). We also assume that each Hamiltonian term depends exclusively either on position or momentum operators, i.e., we write
\begin{equation}
\op{H}(\op{\bm{p}},\op{\bm{x}},t)=\op{H}_1(\op{\bm{p}},t)+\op{h}_2(\op{\bm{x}},t).
\label{th22a}
\end{equation} 
In our notation, $\op{\mathcal{F}}$ and $\op{\mathcal{F}}^{-1}$ represent the Fourier transform (FT) and its inverse (IFT), respectively. Note that by applying the procedure shown in Fig.~\ref{SchemeAa} we start with the wave function $\ket{\psi(t_n)}$ and end up with $\ket{\psi(t_{n+1})}$, both in position representation.

\begin{figure}
  \includegraphics[width=0.8\linewidth]{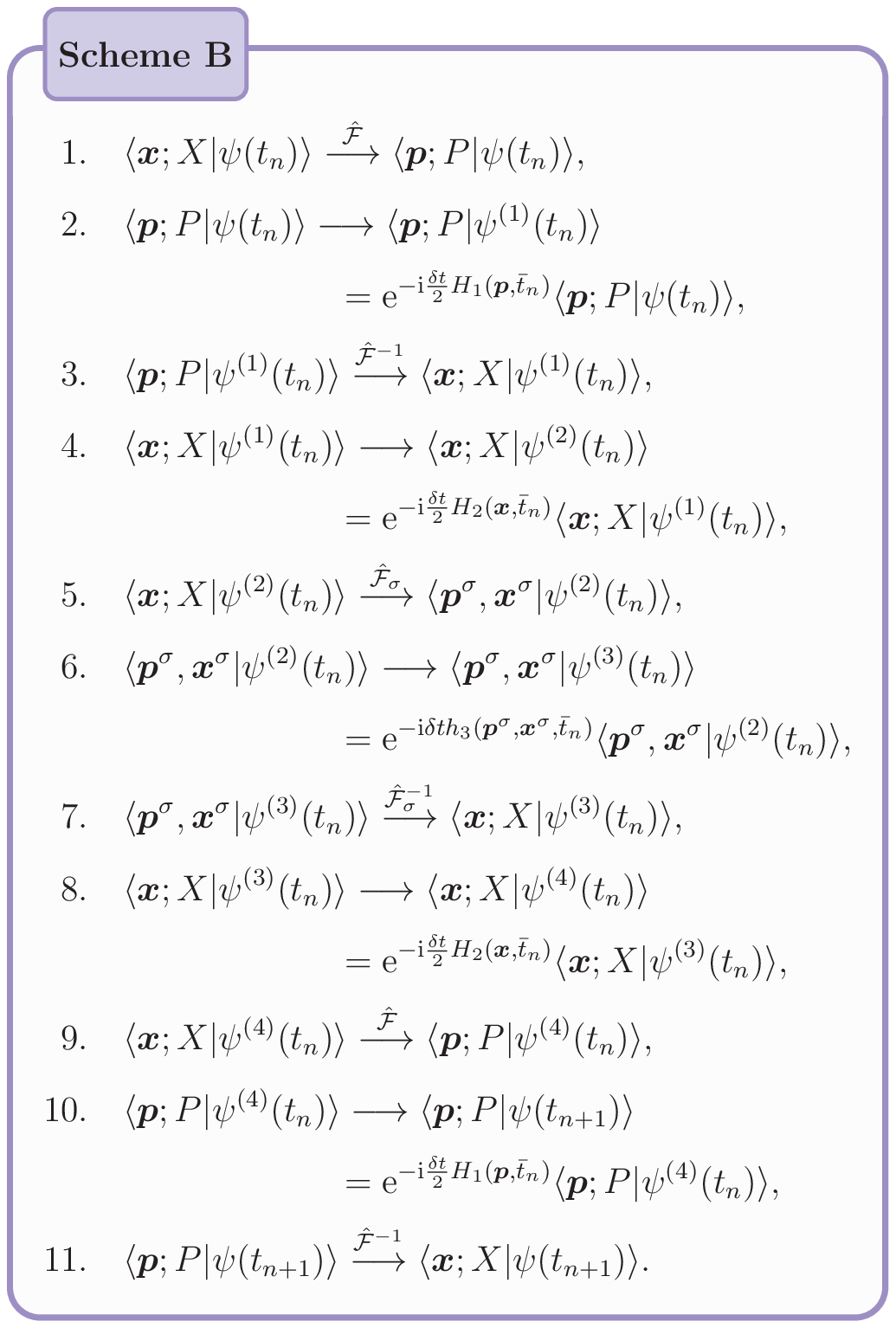}
\caption{Scheme B for the Suzuki-Trotter split-step Fourier method with a Hamiltonian divided into three terms [see, Eq.~\eqref{th15}]. It is assumed 
that the first, $\op{H}_1(\op{\bm{p}},t)$, and the second term, $\op{H}_2(\op{\bm{x}},t)$, depend on the momentum and position 
operators, respectively. However, the third component, $\op{h}_3(\op{\bm{p}}^\sigma,\op{\bm{x}}^\sigma,t)$, contains mixed 
operators and it is presented in the $\sigma$ configuration [see, Eqs.~\eqref{th24} and~\eqref{th25}]. While $\op{\mathcal{F}}$ 
and $\op{\mathcal{F}}^{-1}$ represent FT and IFT, $\op{\mathcal{F}}_\sigma$ and $\op{\mathcal{F}}_{\sigma}^{-1}$ are the partial Fourier transform and its inverse in the $\sigma$ configuration, respectively [Eqs.~\eqref{th27} and~\eqref{th28}]. 
}
\label{SchemeBb}
\end{figure}

The algorithm presented in Scheme A (Fig.~\ref{SchemeAa}) is broadly used in 
strong field physics when applying the dipole approximation. If, however, nondipole effects are going to be studied, this scheme has 
to be suitably modified. This is done by applying additional approximations, the validity of which has to be investigated. To this end, we shall present below an extension of this scheme toward a triply-decomposed Hamiltonian ($\mathcal{K}=3$). Such an extension is based on the 
approximation given by Eqs.~\eqref{th14} and~\eqref{th15}. First, we assume that $\op{H}_1$ and $\op{H}_2$  in Eqs.~\eqref{th10} and~\eqref{th13} 
depend \textit{only} on the momentum $\op{\bm{p}}$ and the position $\op{\bm{x}}$ operators, respectively. However, the third term, $\op{h}_3(\op{\bm{p}},\op{\bm{x}},t)$, depends on both of them. Therefore, the full Hamiltonian $\op{H}(\op{\bm{p}},\op{\bm{x}},t)$ reads
\begin{equation}
\op{H}(\op{\bm{p}},\op{\bm{x}},t)=\op{H}_1(\op{\bm{p}},t)+\op{H}_2(\op{\bm{x}},t)+\op{h}_3(\op{\bm{p}},\op{\bm{x}},t).
\label{th22b}
\end{equation}
It is also assumed that the term with mixed operators, $\op{h}_3(\op{\bm{p}},\op{\bm{x}},t)$, is expressed 
exclusively by commuting components of $\op{\bm{p}}$ and $\op{\bm{x}}$, i.e., it is written in a particular configuration. 
This configuration is defined by a multi-index $\sigma$,
\begin{equation}
\sigma=(\sigma_1,\sigma_2,\dots,\sigma_D),
\label{th23}
\end{equation}
where $\sigma_j=0$ or 1 for $j=1,\dots,D$. Introducing now the operators
\begin{equation}
\op{\bm{x}}^\sigma=((1-\sigma_1)\op{x}_1,\dots,(1-\sigma_D)\op{x}_D)
\label{th24}
\end{equation}
and
\begin{equation}
\op{\bm{p}}^\sigma=(\sigma_1\op{p}_1,\dots,\sigma_D\op{p}_D),
\label{th25}
\end{equation}
we see that all components of $\op{\bm{x}}^\sigma$ and $\op{\bm{p}}^\sigma$ commute with each other. Moreover, the configuration 
$\sigma$ defines the position and momentum vectors ($\bm{x}^\sigma$ and $\bm{p}^\sigma$, respectively) such that their scalar 
product vanishes, $\bm{p}^\sigma\cdot\bm{x}^\sigma=0$. For the purpose of our further analysis, we define also the scalar product in the $\sigma$ configuration as
\begin{equation}
(\bm{p}^\sigma,\bm{x}^\sigma)=\sum_{j=1}^D \sigma_j p_j x_j=\sum_{\sigma_j=1} p_j x_j.
\label{th26}
\end{equation}
This allows us to introduce the partial Fourier transform, denoted by $\op{\mathcal{F}}_\sigma$, such that for any state $\ket{\phi}$ we have
\begin{equation}
\braket{\bm{x};X}{\phi}\stackrel{\op{\mathcal{F}}_\sigma}{\longrightarrow}\braket{\bm{p}^\sigma,\bm{x}^\sigma}{\phi}
\label{th27}
\end{equation}
with
\begin{equation}
\braket{\bm{p}^\sigma,\bm{x}^\sigma}{\phi}=\int\Bigl[\prod_{\sigma_j=1}\dd x_j\Bigr] \ee^{-\ii (\bm{p}^\sigma,\bm{x}^\sigma)}\braket{\bm{x};X}{\phi},
\label{th28}
\end{equation}
which leads to the mixed position-momentum representation of a quantum state $\ket{\phi}$. Similarly to the relations~\eqref{th20} and~\eqref{th21}, 
for the Hamiltonian $\op{h}_3(\op{\bm{p}}^\sigma,\op{\bm{x}}^\sigma,t)$ we arrive at
\begin{equation}
\bra{\bm{p}^\sigma,\bm{x}^\sigma}\ee^{-\ii\delta t\op{h}_3(\op{\bm{p}}^\sigma,\op{\bm{x}}^\sigma,t)}\ket{\phi}
=\ee^{-\ii\delta t{h}_3({\bm{p}}^\sigma,{\bm{x}}^\sigma,t)}\braket{\bm{p}^\sigma,\bm{x}^\sigma}{\phi}.
\label{th29}
\end{equation}
This leads to the Suzuki-Trotter algorithm presented in Scheme B (Fig.~\ref{SchemeBb}). Here, $\op{\mathcal{F}}_\sigma$  
is the partial Fourier transform and $\op{\mathcal{F}}_{\sigma}^{-1}$ is its inverse calculated in the $\sigma$ configuration according to
Eqs.~\eqref{th27} and~\eqref{th28}. In closing this section, we note that this scheme can be easily expanded if the Hamiltonian contains more terms than only one 
in a given configuration. Moreover, a similar approach has been recently used in Refs.~\cite{Lu1,Lu2}.

\section{Physical system}
\label{sec:physical}

Our aim is to investigate the nondipole signatures in ionization by strong nonrelativistic and nearly relativistically intense laser pulses. 
For this purpose, we are going to apply the numerical schemes A and B presented in Sec.~\ref{sec:theory} to solve the one-electron 
Schr\"odinger equation. When solving TDSE numerically, calculations are always limited to finite regions. However, in order to avoid the occurrence 
of unphysical reflection effects from boundaries, this area has to be large enough. This significantly affects the memory being used and the computation 
time. For this reason, the theoretical analysis is often limited to one- or two-dimensional (1D or 2D) spaces. Since nondipole effects (attributed to the presence of the magnetic field) do not occur in 1D spaces, we will investigate ionization of a single-electron atom in two dimensions.

\subsection{Atomic system}
\label{sec:atomic}

Let us consider a 2D hydrogen-like atom with a binding soft-core Coulomb potential, 
\begin{equation}
V(\bm{x})=-\frac{\mathcal{Z}}{\sqrt{x^2+a^2\exp(-x/a)}},
\label{ps1}
\end{equation}
where ${\cal Z}$ is the atomic number, $\bm{x}=(x_1,x_2)$ is the two-dimensional position vector, and 
$x=|\bm{x}|=\sqrt{x_1^2+x_2^2}$ represents its norm. To avoid the presence of quadrupole and higher multipole terms, 
we have introduced the exponential term $a^2\exp(-x/a)$ in the denominator of Eq.~\eqref{ps1} ($a>0$). Note that in our model $V(\bm{x})$ tends asymptotically to the pure Coulomb potential for large $x$,
\begin{equation}
V(\bm{x})=-\frac{\mathcal{Z}}{x}+O\Bigr(\frac{\ee^{-x/a}}{x^3}\Bigl).
\label{ps2}
\end{equation}
Setting $a=1.10609a_0$ and $\mathcal{Z}=1$, the ground state energy in the potential~\eqref{ps1} turns out to be identical as for a 3D hydrogen atom, $E_B=-0.5E_0$. Moreover,
the ground state wave function, $\psi_B(\bm{x})$, is determined either by applying the shooting method for the radial differential equation or 
the Feynman-Kac method for imaginary times. Both approaches give nearly the same results for $E_B$ and $\psi_B(\bm{x})$. Then, the function 
$\psi_B(\bm{x})$ is used as the initial state for the time-propagation of the Schr\"odinger equation~\eqref{th1} with the atomic Hamiltonian
\begin{equation}
\op{H}_{\mathrm{at}}(\op{\bm{p}},\op{\bm{x}})=\frac{1}{2\me}\op{\bm{p}}^2+V(\op{\bm{x}}).
\label{ps3}
\end{equation}
As we have checked, by choosing $\delta t=0.01t_0$ in the Scheme A (Fig.~\ref{SchemeAa}), the state $\psi_B(\bm{x})$ acquires the phase 
factor $\exp(-\ii E_B t)$ for $0<t<100t_0$. Thus, $\psi_B(\bm{x})$ is in fact the eigenstate of the atomic Hamiltonian~\eqref{ps3} 
with eigenvalue $E_B$.

\subsection{Hamiltonian decomposition}
\label{sec:hsplitting}

The electron interaction with the laser field is accounted for by applying the standard minimal coupling prescription,
\begin{equation}
\op{H}(\op{\bm{p}},\op{\bm{x}},t)=\frac{1}{2\me}\bigl(\op{\bm{p}}-e\bm{A}(\op{\bm{x}},t)\bigr)^2+V(\op{\bm{x}}),
\label{ps4}
\end{equation}
where $e$ is the electron charge whereas the vector potential $\bm{A}(\op{\bm{x}},t)$ describes the laser pulse. 
For our numerical illustrations we choose
\begin{equation}
\bm{A}(\op{\bm{x}},t)=(A(t-\op{x}_2/c),0),
\label{ps5}
\end{equation}
which corresponds to a laser pulse propagating along the $x_2$ direction and polarized linearly along the $x_1$ direction. 
Hence, the total Hamiltonian in Eq.~\eqref{ps4} can be written in the form
\begin{equation}
\op{H}(\op{\bm{p}},\op{\bm{x}},t)=\op{H}_1(\op{\bm{p}},t)+\op{H}_2(\op{\bm{x}},t)+\op{h}_3(\op{\bm{p}}^\sigma,\op{\bm{x}}^\sigma,t),
\label{ps6}
\end{equation}
with
\begin{align}
\op{H}_1(\op{\bm{p}},t)=&\frac{1}{2\me}\op{\bm{p}}^2,\nonumber \\ 
\op{H}_2(\op{\bm{x}},t)=&\frac{1}{2\me}\bigl(eA(t-\op{x}_2/c)\bigr)^2+V(\op{\bm{x}}),
\label{ps7}
\end{align}
and
\begin{equation}
\op{h}_3(\op{\bm{p}}^\sigma,\op{\bm{x}}^\sigma,t)=-\frac{e}{\me}A(t-\op{x}_2/c)\cdot\op{p}_1.
\label{ps8}
\end{equation}
This means that, according to our classification, $\op{h}_3$ is in the configuration $\sigma=(1,0)$; thus, the numerical Scheme B (Fig.~\ref{SchemeBb}) can be applied.

In contrast, in the dipole approximation, the operator $A(t-\op{x}_2/c)$ is replaced by the function $A(t)$. In
this case, the Hamiltonian~\eqref{ps4} becomes
\begin{equation}
\op{H}_d(\op{\bm{p}},\op{\bm{x}},t)=\frac{1}{2\me}\Bigl[\bigl( \op{p}_1-eA(t)\bigr)^2+\op{p}_2^2\Bigr]+V(\op{\bm{x}}),
\label{ps9}
\end{equation}
which defines the splitting required for the application of Scheme A. 

To account for nondipole corrections in the lowest order, we make an approximation
\begin{equation}
A(t-\op{x}_2/c)=A(t)+\frac{1}{c}\op{x}_2\mathcal{E}(t)+O(1/c^2),
\label{ps10}
\end{equation}
where $\mathcal{E}(t)=-\dot{A}(t)$ is the electric component of the laser pulse in the dipole approximation. 
Hence, the Hamiltonian~\eqref{ps4}, up to terms proportional to $1/c^2$, becomes
\begin{equation}
\op{H}_{d1}(\op{\bm{p}},\op{\bm{x}},t)=\op{H}'_1(\op{\bm{p}},t)+\op{H}'_2(\op{\bm{x}},t)+\op{h}'_3(\op{\bm{p}}^\sigma,\op{\bm{x}}^\sigma,t),
\label{ps11}
\end{equation}
with
\begin{align}
\op{H}'_1(\op{\bm{p}},t)=&\frac{1}{2\me}\op{\bm{p}}^2,\nonumber \\ 
\op{H}'_2(\op{\bm{x}},t)=&\frac{1}{2\me}\bigl(eA(t)\bigr)^2+\frac{1}{\me c}e^2A(t)\mathcal{E}(t)\op{x}_2+V(\op{\bm{x}}),
\label{ps12}
\end{align}
and
\begin{equation}
\op{h}'_3(\op{\bm{p}}^\sigma,\op{\bm{x}}^\sigma,t)=-\frac{e}{\me}\Bigl(A(t)+\frac{1}{c}\mathcal{E}(t)\op{x}_2\Bigr)\cdot\op{p}_1.
\label{ps13}
\end{equation}
Note that, thanks to the properties of Scheme B, it is not necessary to perform an additional unitary transformation in order to get rid of the term 
$\op{x}_2\op{p}_1$ in Eq.~\eqref{ps13}. Since our analysis is carried out in the inertial reference frame, in which the center of 
the binding potential is at rest and no noninertial forces act on photoelectrons, the interpretation of the results that follow from our numerical 
analysis is relatively simple. Moreover, higher nondipole terms, as well as the relativistic mass corrections, can be easily accounted for without substantial changes in the algorithm shown in Scheme B.

\subsection{Laser pulse}
\label{sec:laser}

In this work we limit our considerations to ionization driven by flat-top pulses which are smoothly turned on 
and off. This allows us, for instance, to study effects that are typically observed for monochromatic plane waves, provided that the pulses 
are long enough. For example, such pulses can be described by a supergaussian envelope of the type $\exp[-(t/\tau)^{N_\mathrm{env}}]$, where $N_\mathrm{env}$ 
is an integer number much larger than 2, as this function is nearly constant for $|t|<\tau$. Thus, we assume that $A (t)$ is of the form
\begin{equation}
A(t)=A_0\exp\Bigl[-\Bigl(\eta\, \frac{\omega t-\pi N_\mathrm{osc}}{\pi N_\mathrm{osc}}\Bigr)^{N_\mathrm{env}}\Bigr]\sin(\omega t+\chi).
\label{ps14}
\end{equation}
Note that the function $A(t)$ relates not only to the electric field strength $\mathcal{E}(t)$, as mentioned before, but also to the 
electron displacement in the laser field $\alpha_D(t)$~\cite{Henneberger,KrollWatson},
\begin{equation}
\alpha_D(t)=-\frac{e}{\me}\int_{-\infty}^t \dd\tau A(\tau).
\label{ps15}
\end{equation}
In our further analysis we choose $\eta=1.35$ and $N_\mathrm{env}=12$ in Eq.~\eqref{ps14}, as for such parameters $A(t)$, $\alpha_D(t)$, and ${\mathcal E}(t)$ 
vanish for $t<0$ and $t>2\pi N_\mathrm{osc}/\omega$ within the accuracy of our numerical calculations. This property is illustrated in Figs.~\ref{lasgrs3osc2a} and \ref{lasgrs3osc50a} below. Moreover, $\omega$ and $\chi$ in Eq.~\eqref{ps14} are the carrier frequency and the carrier envelope phase of the pulse, respectively, and $A_0$ sets up the maximum amplitude of field oscillations.

\section{Numerical details}
\label{sec:numeric}

Before investigating the nondipole signatures in ionization let us first discuss some relevant details of our numerical analysis. 
First of all, the spatial region is defined by the parameter $x_0$ such that $-x_0\leqslant x_1,x_2 < x_0$. The number of points in this domain is fixed by an integer $K$,
\begin{equation}
x_{1,j}=-x_0+(j-1)\Delta x, \, \Delta x=\frac{2x_0}{2^K}, \, j=1,\dots, 2^K,
\label{nd1}
\end{equation}
 and similarly for the second Cartesian coordinate $x_2$. Thus, the discretization in momentum space corresponds to
\begin{equation}
p_{1,j}=-\frac{\pi 2^K}{2x_0}+(j-1)\Delta p,\, \Delta p=\frac{\pi}{x_0},
\label{nd2}
\end{equation}
and the same for $p_2$.

In order to reduce the boundary reflection effects we introduce the mask function $M(x)$ ($x=\sqrt{x_1^2+x_2^2}$), which is given by
\begin{equation}
M(x)=\begin{cases}
1, & x<r_1, \cr
\bigl[\cos\bigl(\frac{\pi}{2}\frac{x-r_1}{r_2-r_1}\bigr)\bigr]^{1/8}, & r_1\leqslant x\leqslant r_2, \cr
0, & x>r_2,
\end{cases}
\label{nd3}
\end{equation}
with $0<r_1<r_2\leqslant x_0$. This function multiplies $\exp(-\ii\delta t h_2(\bm{x},\bar{t}_n))$ in Scheme A (Fig.~\ref{SchemeAa}) as well as
$\exp(-\ii\frac{\delta t}{2} H_2(\bm{x},\bar{t}_n))$ in Scheme B (Fig.~\ref{SchemeBb}). As discussed in Ref.~\cite{Krause1992a}, its role is equivalent 
to introducing an absorbing potential at the boundaries.

The wave functions $\braket{\bm{x};X}{\psi(t)}$, $\braket{\bm{p};P}{\psi(t)}$, and $\braket{\bm{p}^\sigma,\bm{x}^\sigma}{\psi(t)}$ determine the 
probability distributions in the position and momentum spaces, $P(\bm{x},t)$ and $\tilde{P}(\bm{p},t)$, as well as in the mixed representation, 
$\check{P}(\bm{p}^\sigma,\bm{x}^\sigma,t)$, respectively. The normalization of these distributions is chosen such that
\begin{align}
\sum_{j=1}^{2^K}\sum_{\ell=1}^{2^K} P(x_{1,j},x_{2,\ell},t)(\Delta x)^2&=1, \label{nd4}\\
\sum_{j=1}^{2^K}\sum_{\ell=1}^{2^K} \tilde{P}(p_{1,j},p_{2,\ell},t)(\Delta p)^2&=1,
\label{nd5}
\end{align}
and
\begin{equation}
\sum_{j=1}^{2^K}\sum_{\ell=1}^{2^K} \check{P}(p^{\sigma}_{j},x^{\sigma}_{\ell},t)\Delta p\Delta x=1.
\label{nd5a}
\end{equation}
Note, however, that due to the leakage of probability from the finite integration region, the above normalizations are strictly fulfilled only for the initial times.
With this convention the averages of operators $B(\op{\bm{x}},t)$, $C(\op{\bm{p}},t)$, and  $D(\op{\bm{p}}^\sigma,\op{\bm{x}}^\sigma,t)$ are equal to
\begin{align}
\langle B(\op{\bm{x}},t)\rangle=& \bra{\psi(t)}B(\op{\bm{x}},t)\ket{\psi(t)} \\
=&\sum_{j=1}^{2^K}\sum_{\ell=1}^{2^K} B(x_{1,j},x_{2,\ell},t)P(x_{1,j},x_{2,\ell},t)(\Delta x)^2, \label{nd6}\nonumber\\
\langle C(\op{\bm{p}},t)\rangle=& \bra{\psi(t)}C(\op{\bm{p}},t)\ket{\psi(t)} \\
=&\sum_{j=1}^{2^K}\sum_{\ell=1}^{2^K} C(p_{1,j},p_{2,\ell},t)\tilde{P}(p_{1,j},p_{2,\ell},t)(\Delta p)^2, \label{nd7}\nonumber\\
\langle D(\op{\bm{p}}^\sigma,\op{\bm{x}}^\sigma,t)\rangle=& \bra{\psi(t)}D(\op{\bm{p}}^\sigma,\op{\bm{x}}^\sigma,t)\ket{\psi(t)} \\
=&\sum_{j=1}^{2^K}\sum_{\ell=1}^{2^K} D(p^{\sigma}_{j},x^{\sigma}_{\ell},t)\check{P}(p^{\sigma}_{j},x^{\sigma}_{\ell},t)\Delta p\Delta x; \nonumber
\label{nd7a}
\end{align}
thus, accounting for the depletion of probability in the integration region.
In the calculation of the Fast Fourier Transform we use the relevant procedures from the Intel\textsuperscript{\textregistered} Math Kernel Library.

Because of the finite speed of light, we always start the time evolution at $t_\mathrm{i}=-x_0/c$ and end up for times not smaller than $t_\mathrm{f}=2\pi N_\mathrm{osc}/\omega+x_0/c$. Due to this choice the laser pulse has not yet entered the space region for $t<t_\mathrm{i}$ and has already left it entirely for $t>t_\mathrm{f}$.

\section{Mean-value characteristics of electron wave packets}
\label{sec:Mean-value}

\begin{figure*}
  \includegraphics[width=0.8\linewidth]{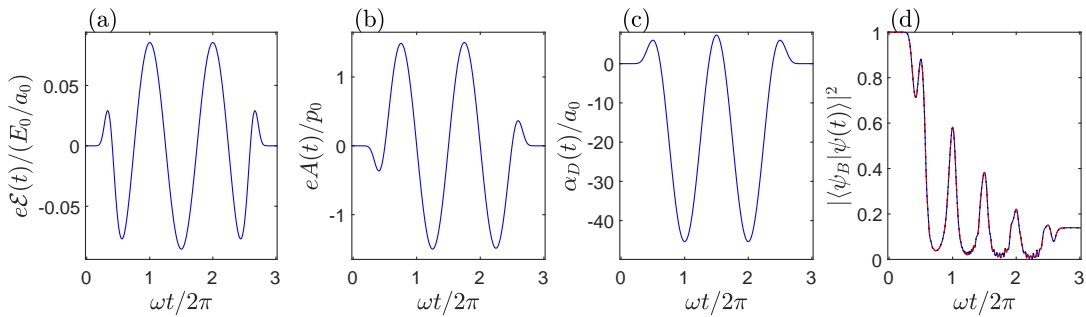}
\caption{The laser pulse functions [(a) electric field strength, (b) vector potential, and (c) displacement function] for $|e|A_0=2p_0$, 
$\omega=0.057E_0$, $N_\mathrm{osc}=3$, $N_\mathrm{env}=12$, $\chi=0$, and $\eta=1.35$. For such parameters we deal with a pulse comprising one cycle 
in the flat-top portion of the envelope. In panel (d) the time-variation of the ground state population is presented for dynamics 
described by the Hamiltonians: \eqref{ps6} [full nondipole calculation, solid blue line], \eqref{ps9} [dipole approximation, dotted black line], 
and \eqref{ps11} [nondipole corrections, dashed red line]. As we see, for these parameters the nondipole effects are rather hardly visible.
}
\label{lasgrs3osc2a}
\end{figure*}

In classical mechanics the well-known signature of nondipole effects in a linearly polarized monochromatic plane wave is the figure-eight
motion of a charged particle. This shape describes a trajectory of a particle in the coordinate system in which it rests on average
as well as it describes the particle acceleration, both in the plane defined by the polarization vector and the direction of wave propagation. 
This property can be derived using the relativistic Hamilton-Jacobi equation~\cite{LandauL2} or the relativistic Newton 
equation with the Lorentz force~\cite{FKK18}.
For the latter, three invariants of motion can be obtained that determine the momentum of a particle. In the coordinate system chosen by us and neglecting the relativistic mass corrections (which are of the order of $1/c^2 $), these invariants lead to the relations,
\begin{equation}
v_1(t)=-\frac{e}{\me}A(t_r), \quad v_2(t)=\frac{1}{2c}[v_1(t)]^2,
\label{cg1}
\end{equation}
where $t_r=t-x_2(t)/c$. Thus, for the acceleration vector we obtain
\begin{equation}
a_1(t)=\dot{v}_1(t), \quad a_2(t)=\frac{v_1(t)}{c}a_1(t).
\label{cg2}
\end{equation}
In particular, if $v_1(t)=v_0\cos(\omega t_r+\phi_0)$ (with an arbitrary phase $\phi_0$), then $a_1(t)=-v_0\omega\sin(\omega t_r+\phi_0)+O(1/c^2)$.
Hence, after squaring the second equation in~\eqref{cg2}, we arrive at the relation,
\begin{equation}
a_2^2=\frac{1}{c^2\omega^2}(v_0^2\omega^2-a_1^2)a_1^2,
\label{cg3}
\end{equation}
which exactly corresponds to the symmetric figure-eight for the acceleration. The above derivation neglects the influence of the binding potential that can lead to some modifications of this picture, as we shall discuss it shortly.

In quantum mechanics analogues of the position, velocity, and acceleration vectors are the mean values of the  corresponding operators. For the position, 
we have
\begin{equation}
\bm{x}(t)=\bra{\psi(t)}\op{\bm{x}}\ket{\psi(t)}.
\label{cg4}
\end{equation}
For the velocity, $\bm{v}(t)=\dot{\bm{x}}(t)$, depending on the choice of the Hamiltonian, we get
\begin{equation}
\bm{v}(t)=
\begin{cases}
\frac{1}{\me}\bra{\psi(t)}\op{\bm{p}}-e\bm{A}(t-\frac{1}{c}\op{\bm{x}}) \ket{\psi(t)}, & \textrm{for \eqref{ps6}} ,\cr
\frac{1}{\me}\bra{\psi(t)}\op{\bm{p}}-e\bm{A}(t)-\frac{e}{c}\bm{\mathcal{E}}(t)\cdot\op{\bm{x}} \ket{\psi(t)}, & \textrm{for \eqref{ps11}} , \cr
\frac{1}{\me}\bra{\psi(t)}\op{\bm{p}}-e\bm{A}(t) \ket{\psi(t)}, & \textrm{for \eqref{ps9}} .
\end{cases}
\label{cg5}
\end{equation}
And finally, for the acceleration we use $\bm{a}(t)=\dot{\bm{v}}(t)$. Note, that in the dipole approximation the components parallel to the propagation direction of all these vectors vanish, i.e., $x_2(t)=0$ and the same for $v_2(t)$ and $a_2(t)$.
Further, let us consider two cases for which the dynamics of the mean values defined above show different behavior.

\subsection{Low-frequency case ($\omega \ll |E_B|$)}
\label{sec:Low}

\begin{figure*}
  \includegraphics[width=0.8\linewidth]{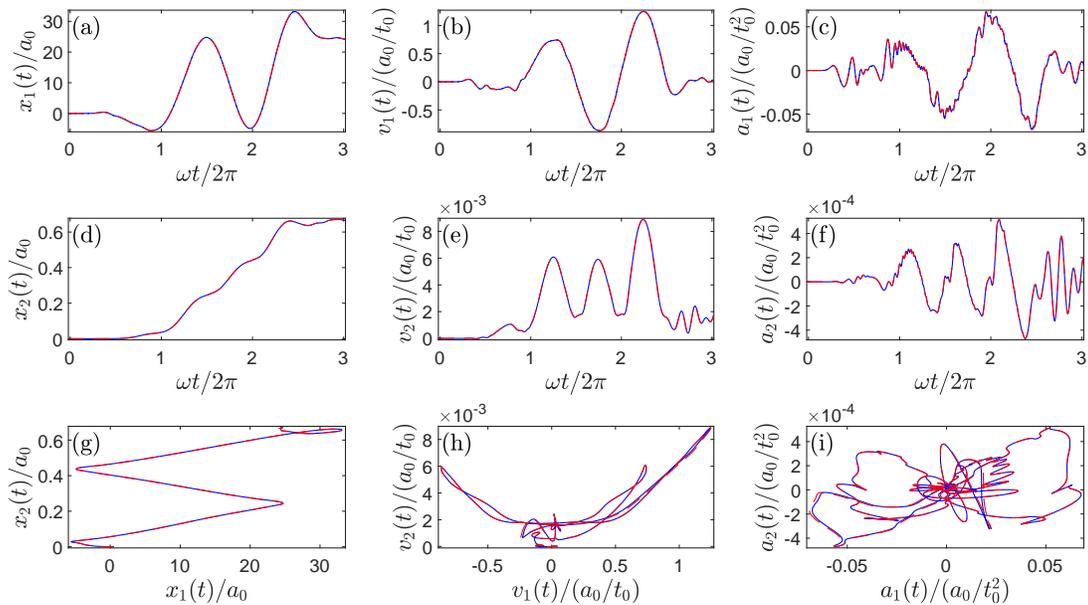}
\caption{In the corresponding panels we present the time-variations of mean positions (a,d), velocities (b,e), and accelerations (c,f) for the laser 
pulse parameters defined in the caption of Fig.~\ref{lasgrs3osc2a}. In panel (g) we demonstrate the relation between the mean position parallel to 
the laser pulse propagation direction, $x_2(t)$, and the mean position perpendicular to it, $x_1(t)$. Panels (h) and (i) show the same relations, 
but for the mean velocities and accelerations, respectively. In panels (a,b,c) we compare the mean values for three Hamiltonians: \eqref{ps6} 
[full nondipole calculation, solid blue line], \eqref{ps9} [dipole approximation, dotted black line], and \eqref{ps11} [nondipole correction, dashed 
red line]. In the remaining panels the comparison is made only for Hamiltonians: \eqref{ps6} [solid blue line] and \eqref{ps11} [dashed red line], 
as in the dipole approximation the mean values parallel to the propagation direction vanish. Our results demonstrate that the lowest 
order nondipole corrections to the full Hamiltonian describe the dynamics of the electron wave packet sufficiently well.
}
\label{3osc12kt01x300a2all}
\end{figure*}

\begin{figure*}
  \includegraphics[width=0.8\linewidth]{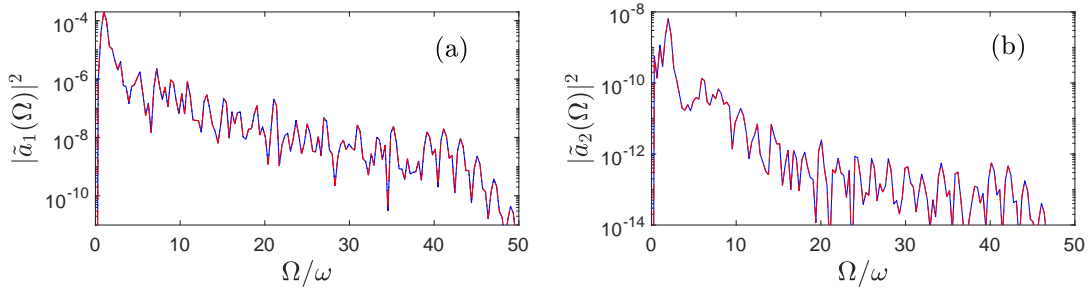}
\caption{Fourier transform modulus squared of the mean acceleration components which are either parallel: (a)  
to the polarization vector or (b) to the pulse propagation direction.
The acceleration components are calculated for different Hamiltonians, specified by Eq.~\eqref{ps6} [full nondipole calculation, 
solid blue line], Eq.~\eqref{ps9} [dipole approximation, dotted black line], and Eq.~\eqref{ps11} [nondipole correction, dashed red line], 
by numerically differentiating the mean velocity vector~\eqref{cg5}. The laser field parameters are defined in 
Fig.~\ref{lasgrs3osc2a}. 
}
\label{hhg3osc12kt01x300a2}
\end{figure*}

First, we consider the laser pulse of frequency $\omega=0.057E_0$ that corresponds roughly to the wavelength of $800$~nm. 
In Fig.~\ref{lasgrs3osc2a} we present the electric field strength, vector potential, and the displacement for the three cycle pulse with the supergauss 
envelope and $|eA_0|=1.5p_0$. For these parameters, the ponderomotive energy equals $U_p=e^2A_0^2/4\me=0.5625E_0$ and the 
corresponding Keldysh parameter is $\gamma=\sqrt{|E_B|/2U_p}=2/3$, i.e., we consider the intermediate multiphoton-tunneling regime. 
In Fig.~\ref{lasgrs3osc2a}(d) we demonstrate the population of the ground state during the time evolution, which exhibits peaks when the modulus 
of the electric field strength is maximum. It is well known that, if such revivals take place, one can expect to observe the high-order harmonics in the spectrum of emitted radiation \cite{Schafer1993,KulanderClassic,Corkum1993,Lewenstein1994}, as it is going to be discussed shortly.

In Fig.~\ref{3osc12kt01x300a2all}, we present the mean values of the electron position, velocity, and acceleration. 
All these quantities in the direction of the laser pulse polarization are nearly independent of the choice of Hamiltonians considered in Sec.~\ref{sec:hsplitting}. 
The discrepancies appear if we consider components parallel to the pulse propagation direction, which vanish in the dipole approximation. 
However, if nondipole terms in the Hamiltonian are accounted for we observe the effects related to radiation pressure that push the center of electron 
wave packet in the light propagation direction. The electron trajectory, Fig.~\ref{3osc12kt01x300a2all}(g), if calculated in the reference frame of 
vanishing averaged velocity, resembles the figure eight, whereas the velocity dependence, Fig.~\ref{3osc12kt01x300a2all}(h), is quite well described by the parabola \eqref{cg1}. The discrepancies are related to the interaction (rescattering) of electron with the binding potential in the laser pulse, which is not accounted for in the classical analysis discussed above.

Even greater differences with the classical picture are manifested in the time dependence of electron acceleration, shown in Fig.~\ref{3osc12kt01x300a2all}(h). 
Here, in both cases, we observe very fast changes caused by rescattering processes. The consequence of these changes is 
that many Fourier coefficients do not disappear, which leads to the formation of a wide spectrum of high-order harmonics, as presented in 
Fig.~\ref{hhg3osc12kt01x300a2}. There we observe the broad distributions of the Fourier transform modulus squared of both acceleration 
components which, according to the Larmor formula, are proportional to the intensity of generated radiation. The cut-offs of both plateaux are given 
by the well-known formula~\cite{Krause1992b,JKPbook},
\begin{equation}
\frac{\Omega_{\mathrm{cut-off}}}{\omega}=\frac{1}{\omega}(|E_B|+3.17U_p)\approx 40.
\label{cg6}
\end{equation}
The nondipole signature in the high-order harmonic generation is that harmonics are also emitted in the direction of light polarization. 
Moreover, contrary to the `ordinary' harmonics [Fig.~\ref{hhg3osc12kt01x300a2}(a)], for which the spectrum consists mainly of odd harmonics, for the 
spectrum generated due to nondipole effects [Fig.~\ref{hhg3osc12kt01x300a2}(b)] we observe mostly even harmonics. A qualitative explanation of this fact 
can be based on the quadratic dependance of $v_2(t)$ on $v_1(t)$ [see Eq.~\eqref{cg1} and its consequence, Eq.~\eqref{cg2}]. Numerical analysis of $v_1(t)$ 
shows that in its Fourier decomposition the modulus squared of the constant term is around one order of magnitude smaller than 
the first harmonic for $\Omega=\omega$. Hence, since the odd harmonics dominate in the Fourier decomposition of $v_1(t)$ and $a_1(t)$, in the corresponding 
decomposition of $a_2(t)$ we mostly observe even harmonics.

In conclusion, for low-frequency laser pulses for which the Keldysh parameter $\gamma$ is smaller than unity (in fact, this is the most interesting case 
in the context of high-order harmonic generation), nondipole effects are hardly visible for as long as those pulses contain only few
oscillations. On the other hand, if the number of oscillations is large, the nondipole signatures are amplified at the end of the pulse, when the mean position of the 
electron wave packet [i.e., $x_2(t)$] is far from the potential center and the dipole approximation~\eqref{ps9} is no longer valid. 
However, this case is difficult to analyze numerically because during the time evolution a significant portion of the electron wave packet leaks from 
the integration region.

\begin{figure*}
  \includegraphics[width=0.8\linewidth]{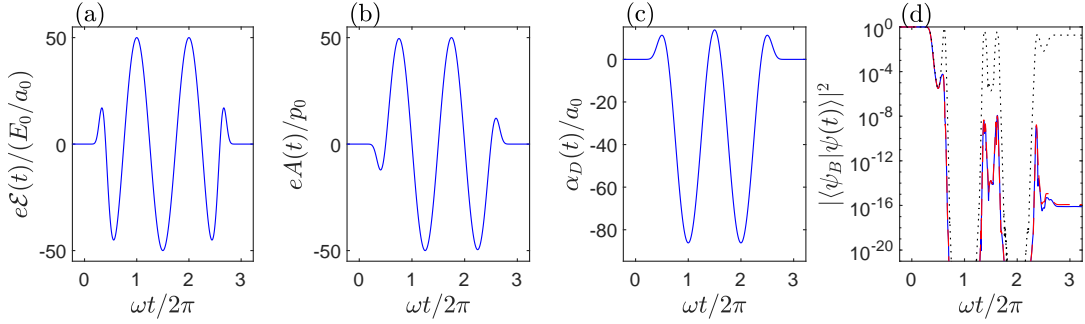}
\caption{The laser pulse functions [(a) electric field strength, (b) vector potential, and (c) displacement function] for $|e|A_0=50p_0$, $\omega=E_0$, 
$N_\mathrm{osc}=3$, $N_\mathrm{env}=12$, $\chi=0$, and $\eta=1.35$. For such parameters we deal with a pulse comprising one cycle in the flat-top 
portion of the envelope. In the panel (d) the time-variation of the ground state population is presented for dynamics described 
by the Hamiltonians: \eqref{ps6} [full nondipole calculation, solid blue line], \eqref{ps9} [dipole approximation, dotted black line], and~\eqref{ps11} 
[nondipole corrections, dashed red line]. For such pulse parameters, significant nondipole effects are observed.
}
\label{lasgrs3osc50a}
\end{figure*}

\begin{figure*}
  \includegraphics[width=0.8\linewidth]{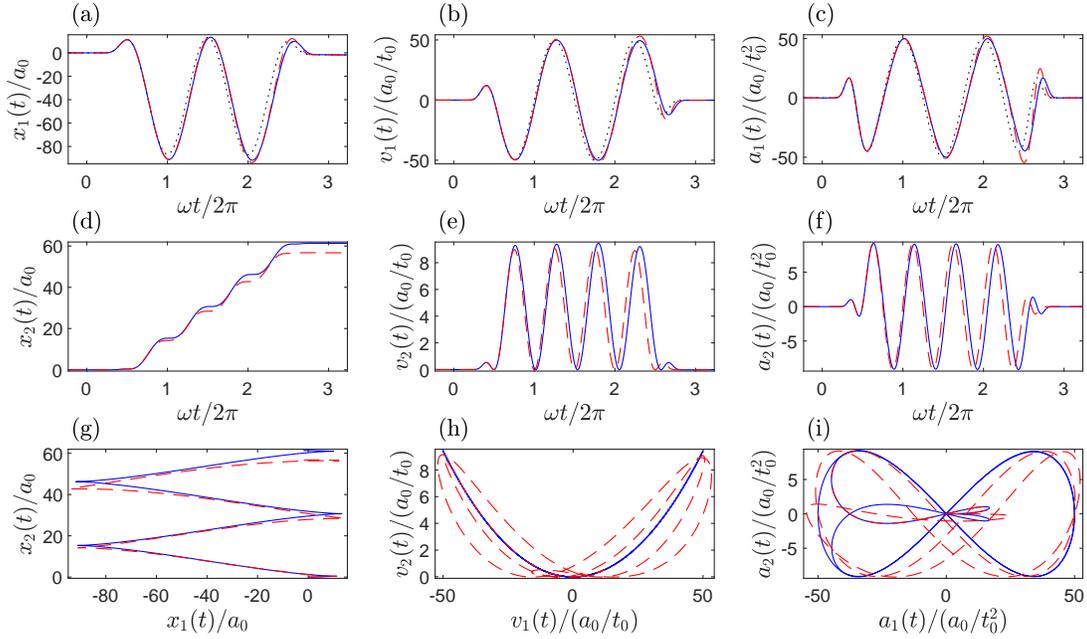}
\caption{The same as in Fig.~\ref{3osc12kt01x300a2all}, but for the laser pulse parameters defined in the caption of Fig.~\ref{lasgrs3osc50a}. 
For these parameters we observe discrepancies between the full Hamiltonian~\eqref{ps6} (solid blue line) and its lowest order nondipole 
approximation~\eqref{ps11} (dashed red line). For the full Hamiltonian, the averaged values follow very closely the classical pattern.
}
\label{3osc12kt01x200a50all}
\end{figure*}

\subsection{High-frequency case ($\omega > |E_B|$)}
\label{sec:High}

The inspection of the leading order nondipole approximation \eqref{ps10} shows that its validity is limited to
\begin{equation}
\frac{\omega}{c} |x_2(t)| \ll 1.
\label{cg7}
\end{equation}
This condition is violated for high-frequency pulses if their intensity and/or the number of cycles are sufficiently large, as under such 
conditions the average position $x_2(t)$ can reach a significant value at the end of the pulse. For this reason, we consider 
the laser pulse of frequency $\omega=E_0$ for which $|e|A_0=50p_0$. Hence, the ponderomotive energy equals $U_p=625E_0$
and the Keldysh parameter is much less than one. This means that we investigate the over-barrier ionization regime in which the atom is immediately ionized
by the laser field. In Fig.~\ref{lasgrs3osc50a} we present: (a) the electric field strength, (b) the vector potential, and (c) the 
displacement function for a three-cycle pulse. In Fig.~\ref{lasgrs3osc50a}(d), the population of the ground state of a 2D-hydrogen
atom during the interaction with the laser field is shown. As we see, if the nondipole corrections are accounted for, the contribution of the ground 
state to the electron wave packet is significantly depleted just in the beginning of the pulse. In addition, the ground state is only 
slightly repopulated when the displacement vanishes.

In order to gain a deeper insight into the time-evolution of the electron wave packet within the laser pulse we shall investigate 
now the mean values. In the first row in Fig.~\ref{3osc12kt01x200a50all} we present the mean position, velocity, and acceleration in the direction 
of light polarization. By comparing these panels with the corresponding panels of Fig.~\ref{lasgrs3osc50a} we conclude that, with a very good agreement,
\begin{equation}
x_1(t)=\alpha_D(t),\, v_1(t)=-\frac{e}{\me}A(t), \, a_1(t)=\frac{e}{\me}\mathcal{E}(t).
\label{cg8}
\end{equation}
This suggests that the time evolution of the electron wave packet is mainly determined by the interaction with the laser field. 
Hence, for the current laser pulse parameters, the rescattering with the residual ion and the Coulomb focusing effects marginally influence the mean values. The same happens also in the dipole approximation, for which the perpendicular shift of the electron wave packet is zero. For all approximations considered in this paper we get nearly the same dependence on time, with the gradually increasing discrepancies as the pulse terminates. Moreover, for longer pulses these differences become even more significant.

The mean values of electron position, velocity, and accelarations calculated along the direction of pulse propagation are presented in the middle row in 
Fig.~\ref{3osc12kt01x200a50all}. This is disregarding the dipole approximation, for which all these components vanish. Here, 
we clearly see the differences between the time evolution determined either by the full Hamiltonian~\eqref{ps6} or by its lowest order nondipole 
approximation \eqref{ps11}. This is of course due to the large radiation pressure that moves the electron wave packet in the direction of light 
propagation to such an extent that the validity condition~\eqref{cg7} is no longer applicable. The breakdown of the lowest order nondipole 
approximation is clearly seen in Figs.~\ref{3osc12kt01x200a50all}(h) and (i). We learn from them that for the full Hamiltonian~\eqref{ps6}
the velocity $v_2(t)$ depends quadratically on $v_1(t)$ and for times corresponding to the flat portion of the envelope the  
acceleration curve adopts nearly a perfect figure-eight shape, as follows from the classical analysis in which the interaction with the binding 
potential is neglected. This is another indication that just after ionization the electron wave packet moves predominantly under the action of the laser 
field, with apparently small corrections resulting from the interaction with the residual ion. This means that, for the current laser field parameters, 
the effects related to rescattering and Coulomb focusing can be disregarded while the electron is still interacting with the laser pulse.

\section{Electron probability distributions}
\label{sec:smd}

\begin{figure*}
  \includegraphics[width=0.89\linewidth]{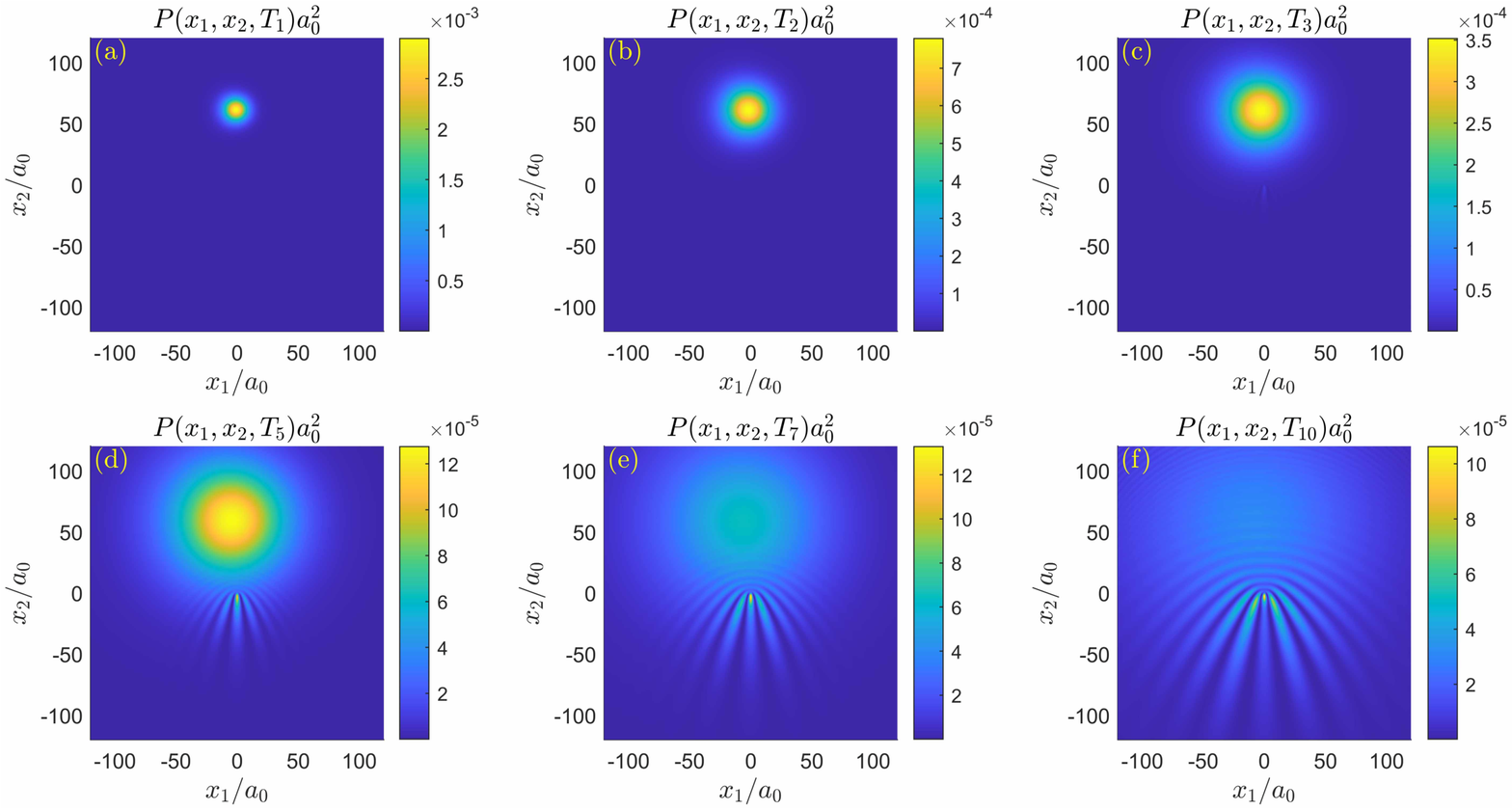}
\caption{Snapshots of the electron probability distribution in position space for selected times 
$T_j=2\pi jN_{\mathrm{osc}}/\omega+x_0/c$, for which the electron wave packet interacts only with the static binding potential. 
The numerical analysis has been carried out for $x_0=200a_0$, $\delta t=0.01t_0$, $K=12$, and for the mask function parameters $r_1=180a_0$ 
and $r_1=199a_0$, as defined in Sec.~\ref{sec:numeric}. The laser pulse parameters are the same as in Fig.~\ref{lasgrs3osc50a}.  
The total probability for the electron wave packet for staying within the integration region is equal to unity 
for $T_{1}$ and $0.88$ for the largest time $T_{10}$.
}
\label{xddistlin}
\end{figure*}
\begin{figure*}
  \includegraphics[width=0.89\linewidth]{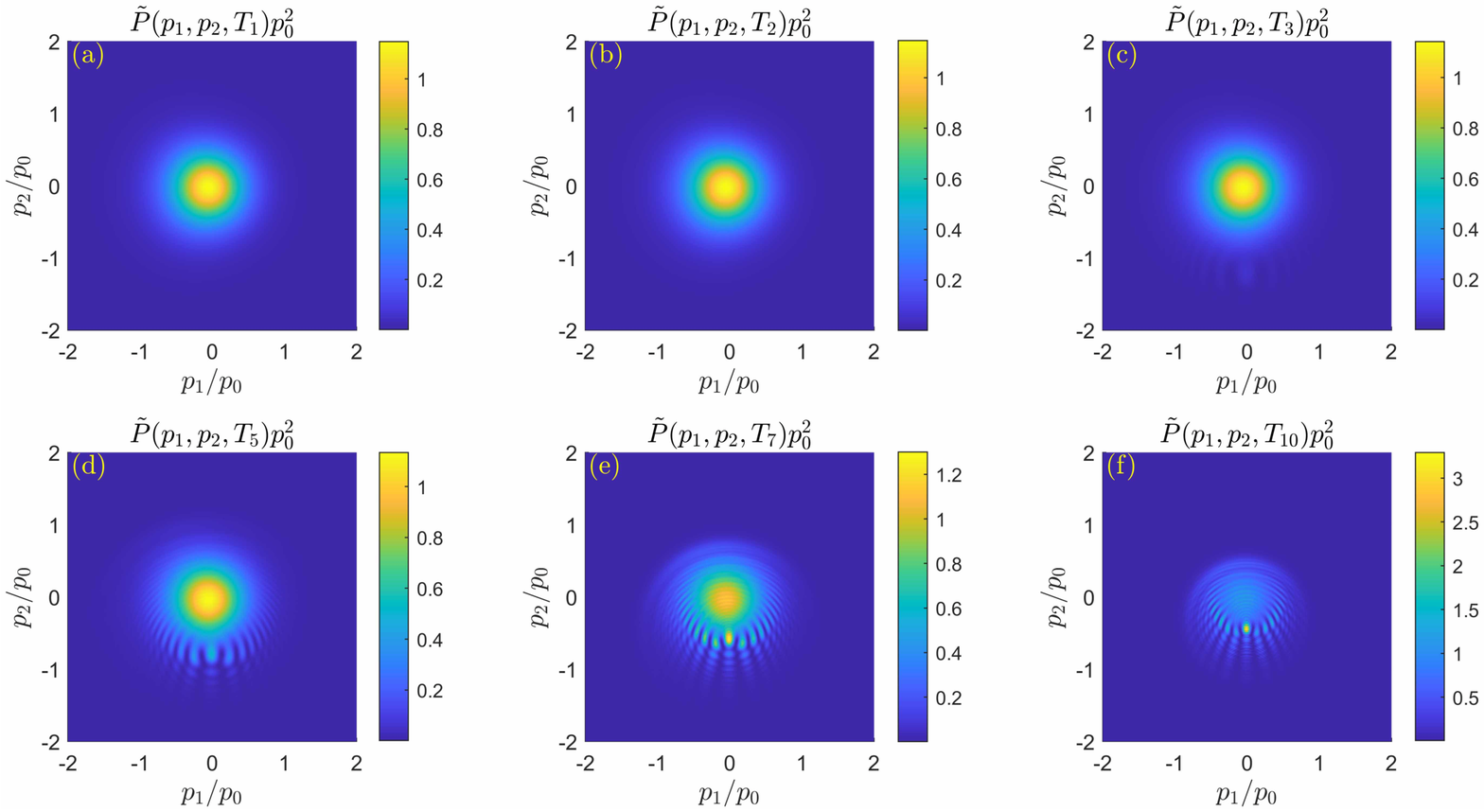}
\caption{The same is in Fig.~\ref{xddistlin} but in the momentum space. The local maxima of these distributions are observed for $p_1=0$ and for negative values of $p_2$ equal to: $p_2=-1.2p_0$ for $T_3$, $p_2=-0.8p_0$ for $T_5$, $p_2=-0.6p_0$ for $T_7$, and $p_2=-0.44p_0$ for $T_{10}$. Starting from $T_7$ these are also the global maxima.
}
\label{pddistlin}
\end{figure*}

For the high-frequency case analyzed in Sec.~\ref{sec:High} we have chosen the laser pulse parameters similar to the ones considered in
Ref.~\cite{Forre2006}. As it follows from there, contrary to the expectations based on the radiation pressure phenomenon, a large portion of electrons 
is emitted opposite to the field propagation direction. Even more, the smaller the photoelectron energy is the more probable emission 
in the opposite direction is expected. This puzzle was initially explained by considering the classical dynamics of electrons in the laser pulse and showing that 
the rescattering generates electron trajectories with final momenta opposite to the pulse propagation direction. This seminal work 
spurred further investigations of the nondipole signatures in ionization, both experimental and theoretical. In particular, it was shown that similar 
effects can be expected for the low-frequency fields. An expanded discussion of experimental works and theoretical models used in this context can be 
found in Ref.~\cite{Haram2020}.

The analysis presented in Sec.~\ref{sec:Low} shows that rescattering and Coulomb focussing in the laser field can explain the 
existence of electrons with negative momenta when ionized by low-frequency pulses. However, for the 
high-frequency fields this interpretation is in contradiction with what we have discussed in Sec.~\ref{sec:High}. Therefore, yet another mechanism 
for the generation of photoelectrons of negative momenta (especially the low-energy ones) has to be proposed.

Note that the electron wave packet spreading significantly affects properties of various strong field processes. 
This is due to the fact that many of them, like HHG~\cite{KulanderClassic,Corkum1993} or generation of high-energy structures in the spectrum of emitted electrons~\cite{MPBB2006}, are interpreted as the result of recombination or rescattering
of electrons returning to the Coulomb potential center. As long as trajectories of these electrons are determined only by their interaction with 
the laser field and as long as scattering is treated in the Born approximation, this picture successfully describes the aforementioned phenomena. Problems arise,
however, when an attempt is made to include the electron interaction with the Coulomb field of the parent ion, e.g. in the eikonal 
approximation. Then, in the theoretical description, certain singularities appear that do not occur in other approaches, 
for example in the Born approximation or in the numerical solution of the Schr\"odinger equation. As shown in Refs.~\cite{FKK2015,KFK19}, 
these singularities arise from neglecting the electron wave packet spreading during its interaction with the laser field
which, on the other hand, is  automatically accounted for in the Born series or in the generalized eikonal approximation~\cite{KFK19}.

As discussed in Sec.~\ref{sec:High}, no signatures (or marginally small signatures) of returning electrons are observed during 
their interaction with the high-frequency laser pulse. On the other hand, the electron wave packet also spreads after the interaction with the laser 
pulse is over. The aim of this section is to show that such a `\textit{post-pulse electron wave packet spreading}' together with its interaction with the parent 
ion lead to the formation of low-energy structures that are shifted opposite
to the laser field propagation direction.

In Figs.~\ref{xddistlin} and~\ref{pddistlin}, we present the snapshots of the post-pulse electron wave packet distributions in the position and 
momentum spaces, respectively, for selected times $T_j=2\pi jN_{\mathrm{osc}}/\omega+x_0/c$. When the interaction with 
the laser field is over, the electron wave packet is moved far from the ionic center, with very small signatures 
indicating the electron scattering by the binding potential [see, Fig.~\ref{xddistlin}(a)]. Both position and momentum distributions are axially 
symmetric (exactly as the initial ground state wave function) with the momentum distribution centered at the origin [cf., Fig.~\ref{pddistlin}(a)]. 
From this moment on, the wave packet begins to spread, but so that its momentum distribution remains unchanged [cf., Figs.~\ref{xddistlin}(b) 
and~\ref{pddistlin}(b)]. We observe such a situation until the Coulomb field begins to noticeably affect the electron quantum dynamics. It is at this point that an interference structure in the momentum distribution of photoelectrons starts to develop due 
to the Coulomb focusing. As a result, a tiny local maximum is formed away from the potential center and, surprisingly, opposite 
to the radiation pressure action [cf., Figs.~\ref{xddistlin}(d) and~\ref{pddistlin}(d)]. With time, this feature gets only enhanced, resulting 
eventually in a pronounced maximum located at negative momenta [cf., Figs.~\ref{pddistlin}(e) and~\ref{pddistlin}(f)]. Hence, we conclude that 
for high-frequency laser pulses, the main reasons for the backward (as compared to the pulse propagation direction) shift of the electron momentum distribution is the post-pulse 
spreading of its wave packet and the Coulomb focusing.

\section{Conclusions}
\label{sec:Conclusions}

We have introduced here the Suzuki-Trotter split-step Fourier method to solve numerically the time-dependent Schr\"odinger equation. It was 
shown that nondipole effects can be efficiently treated within this scheme. This was illustrated while studying 
HHG and strong-field photoionization by intense and finite laser pulses. According to our explorations, for low-frequency pulses the nondipole effects 
in HHG manifest themselves as radiation emitted perpendicular to the laser field propagation direction, which is characterized by the presence of 
predominantly even harmonics. Furthermore, in ionization driven by high-frequency laser pulses, the radiation pressure causes a significant forward drift of the electron wave packet with negligibly small signatures related to its interaction with the binding potential (i.e., we do not observe rescattering or Coulomb focusing effects in the light field). 

A surprising effect in photoionization, i.e., a backward (with respect to the pulse propagation direction) photoelectron drift, has been largely 
discussed (see e.g. Refs.~\cite{Forre2006,Ludwig2014,Maurer2018}). It was found to be caused by a modification of the electron trajectories due to 
Coulomb focusing. Our numerical analysis shows that, at least for the high-frequency pulses, in the description of electron dynamics one has to take 
into account its post-pulse wave packet spreading, which is an inherently quantum phenomenon. In fact, our analysis suggests that the post-pulse dynamics in the Coulomb field is the main cause of the electron backward shift, instead of an in-pulse trajectory modification. This interpretation is compatible with the interference structures for negative momenta. If this mechanism also contributes to the creation of the low-energy photoelectron structures for low-frequency pulses still remains an open question.

\section*{Acknowledgements}
This work is supported by the National Science Centre (Poland) under Grant 
Nos. 2018/30/Q/ST2/00236 (M.C.S and K.K.) and 2018/31/B/ST2/01251 (F.C.V. and J.Z.K.).

\end{document}